\documentclass[useAMS,usenatbib,usegraphicx]{mn2e}
\usepackage{amssymb,amsmath,aas_macros,pifont,hyperref,amsbsy,epsfig,natbib,amsfonts} 

\voffset =- .6in 

\title[Turbulence in the circum-galactic medium]{Turbulence driven by
  structure formation in the circum-galactic medium} 

\author[L. Iapichino et al.]{L. Iapichino$^{1}$\thanks{E-mail:
    luigi@uni-heidelberg.de}, M. Viel$^{2,3}$ and S. Borgani$^{2,3,4}$\\
$^{1}$Universit\"at Heidelberg, Zentrum f\"ur Astronomie, 
Institut f\"ur Theoretische Astrophysik, Philosophenweg 12, D-69120 \\
Heidelberg, Germany\\
$^{2}$INAF, Osservatorio Astronomico di Trieste, via Tiepolo 11, I-34131 Trieste, Italy\\
$^{3}$INFN, Istituto Nazionale di Fisica Nucleare, Via Valerio 2, I-34127 Trieste, Italy\\
$^{4}$Dipartimento di Fisica dell'Universit\`a di Trieste, Sezione di Astronomia, via Tiepolo 11, I-34131 Trieste, Italy\\
}

\begin{document}

\date{Accepted 2013 April 10.  Received 2013 April 10; in original form 2012 October 2}

\pagerange{\pageref{firstpage}--\pageref{lastpage}} \pubyear{2012}

\maketitle

\label{firstpage}

\begin{abstract}
The injection of turbulence in the circum-galactic medium at redshift
$z = 2$ is investigated using the mesh-based hydrodynamic code {\sc
  enzo} and a subgrid-scale (SGS) model for unresolved
turbulence. Radiative cooling and heating by a uniform Ultraviolet
(UV) background are included in our runs and compared with the effect
of turbulence modelling. Mechanisms of gas exchange between galaxies
and the surrounding medium, as well as metal enrichment, are
not taken into account, and turbulence is here driven solely by structure formation
(mergers and shocks). We find that turbulence, both at resolved and
SGS scales, impacts mostly the warm-hot intergalactic medium (WHIM),
with temperature between $10^5$ and $10^7\ {\rm K}$, mainly located
around collapsed and shock heated structures, and in
filaments. Typical values of the ratio of turbulent to thermal
pressure is 0.1 in the WHIM, corresponding to a volume-weighted
average of the SGS turbulent to thermal Doppler broadening $b_{\rm t} / b_{\rm therm} = 0.26$, on length scales below the grid resolution of $25\ \rm{kpc\ h^{-1}}$. 
In the diffuse intergalactic medium (IGM), defined in a range of baryon overdensity $\delta$ between 1 and 50, the importance of turbulence is smaller, but grows as a function of gas density, and the Doppler broadening ratio is fitted by the function $b_{\rm t} / b_{\rm therm} = 0.023 \times \delta^{0.58}$.
\end{abstract}


\section{Introduction}
\label{intro}

The properties of the diffuse gas component, not yet accreted in
collapsed cosmic structures, are subject of study in order to
understand the census of the baryonic matter which does not belong to
galaxies. Early numerical investigations \citep{co99,dco01} have shown
that a substantial fraction (30 to 40 per cent at redshift $z = 0$) of
this gas resides in a baryon phase with temperatures between $10^5$
and $10^7\ {\rm K}$, the Warm-Hot Intergalactic Medium (WHIM), and a
similar amount is contained in a colder ($T < 10^5\ {\rm K}$) diffuse
phase (e.g.~\citealt{tbv10}).

Unfortunately, the physical properties of both phases make their
observational study challenging. Customary diagnostics are Ly$\alpha$
\citep{rauch98} and metal absorption lines at UV and X--ray energies in the spectra of
high-redshift quasars \citep[e.g.][]{richter08}. These tools can be fully exploited only if one
understands the interplay between galaxies and their physics, on the
one side, and the baryonic matter surrounding them on the other side,
both in theory and observations.

Ram pressure stripping, galactic winds, AGN outflows and galaxy-galaxy interactions contribute to the gas transfer between a galaxy and its ambient medium (see \citealt{sd08}, for a review). One important and widely studied facet of this problem is the chemical enrichment of the intra-cluster medium (ICM; \citealt{bft08}). Instead, in this work we will focus on a different but somewhat related point, namely the injection of turbulence in the gas surrounding galaxy-sized haloes. 

The discussion about turbulent motions in the gas at galactic scales has been addressed by several authors. \citet{rsb01} studied the C {\sc iv} absorption in spectra of double quasars at high redshift ($z \geq 2$), thus probing velocity differences (likely to be associated with turbulent motions) on length scales from a few parsecs to a few tens of kiloparsecs. With this technique, they detect a velocity shear on scales larger than a few hundred parsecs, up to $70\ {\rm km\ s^{-1}}$ at a few kiloparsecs. In the same work a complementary approach, based on turbulence motions along the line of sight, derived by the width of absorption-line profiles of the individual C {\sc iv} systems (cf.~also \citealt{rsw96}), gives a root-mean-square (henceforth rms) velocity contribution of $4.7\ {\rm km\ s^{-1}}$ on scales of $300\ {\rm pc}$.

On the numerical side, \citet{od09} in their SPH simulations were able to fit the observed properties of O {\sc vi} absorption lines only by adding in post-processing sub-resolution turbulence, at the level of a few tens of ${\rm km\ s^{-1}}$. Turbulence injected by galactic outflows in the intergalactic medium (henceforth IGM) has been studied also by \citet{ef11}, using a merger tree as input for a spectral turbulence energy model. In that work, the average turbulent Doppler parameter (to be defined in Section \ref{sgs-properties}) peaks at $z \approx 1$ to values of about $1.5\ {\rm km\ s^{-1}}$, with maximum of $25\ {\rm km\ s^{-1}}$. 

In this paper we will study the properties of turbulence in the medium around galaxy-sized haloes by means of grid-based hydrodynamical simulations. With respect to existing, similar studies in this field, we will make use of a subgrid scale (SGS) model for unresolved turbulence. This model, first developed by \citet{snh06} and implemented in an Adaptive Mesh Refinement (AMR) code by \citet{mis09}, has been already used for studying the WHIM properties at cluster scales by \citet{isn11}. The results of that work support the idea that turbulence is driven by different mechanisms in the WHIM and in the ICM, namely by turbulence production at shocks in the former, and by merger processes in the latter. At cluster scales, the level of turbulence in the WHIM grows with time and saturates around $z = 0$.

\citet{isn11} and the present work explore very different length
scales, and together with the scale also the involved physics is
rather different. Earlier in this Introduction, we already mentioned
the interaction processes between galaxies and the surrounding medium;
these processes are important not only for the gas transfer and the
metal enrichment of the circum-galactic medium (hereafter CGM), but also as stirring
agents in the medium. Turbulence around galaxies is expected to be
produced by supernova-driven galactic winds (e.g., \citealt{ef11})
and/or by mergers \citep{rsb01}. Furthermore, the enrichment of the
IGM with metals at low density can be explained if feedback either in
the form of galactic winds or AGN feedback is present: these
mechanisms must be invoked in order to provide relatively good
agreement with observational properties related to quasi-stellar object (QSO) absorption lines statistics \citep{meiksin09}.

In our simulations we make two (over-)simplifying assumptions: only stirring driven by structure formation, i.e.~by mergers and shocks, is considered, and the metal evolution of the gas is not treated. These two assumptions are closely linked to each other: in our turbulence framework, any model of galactic outflow should consist not only of a chemical enrichment model, but also of a model for injecting turbulence energy. This further step, which gathers turbulence, metal enrichment and galactic processes in a single comprehensive model, will be subject for future work. Only in Section \ref{feedback} we perform a preliminary study of the role of star formation, thermal feedback and metal enrichment on the production of turbulence in the CGM; however, already at the present stage of the project, we are able to show the potential of turbulence modelling in this problem, and try to assess what is the relative role of stirring by structure formation. Moreover, we focus mostly on the gas with moderate overdensity 
because it probes the bulk of the IGM at high redshift
(e.g.~\citealt{meiksin09}) and can be studied by means of quasar
absorption lines spectroscopy. Furthermore, the high redshift IGM can
also be used to study the galaxy/IGM interplay by cross-correlation
studies of galaxy properties and absorption lines along line-of-sights
to distant QSOs or galaxies (e.g.~\citealt{steidel10}).  

This paper is structured as follows: in Section \ref{tools} we introduce the numerical techniques, the simulations and the gas phases that will be mainly analysed. In Section \ref{results} the results are presented, in terms of general properties of the runs (Section \ref{general}), subgrid turbulence in the different gas phases (Section \ref{sgs-properties}), in the vicinity of massive haloes (Section \ref{haloes}) and using resolved velocity diagnostics (Section \ref{los}). The results are finally discussed and summarised in Section \ref{discussion}.

\section{Numerical tools}
\label{tools}
The simulations performed in this work were run using the grid-based
hybrid (N-body plus hydrodynamical) code {\sc enzo} (v1.0)
\citep{obb05}\footnote{{\sc enzo} current homepage:
  http://enzo-project.org/}. We simulated the evolution of a
computational box with the comoving size of $10\ \rmn{Mpc}\ h^{-1}$ on
a side, starting from the initial redshift $z = 99$ (with the transfer functions by \citealt{eh99}), to $z = 2$. Our
cosmological parameters are, according to the five-year WMAP results
\citep{kdn09}, $\Omega_\rmn{\Lambda} = 0.721$, $\Omega_\rmn{m} =
0.279$, $\Omega_\rmn{b} = 0.046$, $h = 0.7$, $\sigma_8 = 0.817$, and
$n=0.96$.

The computational box is resolved with a root grid of $400^3$ cells,
and the same number of N-body DM particles (each of them with a mass
of $1.01 \times 10^6\ \rmn{M_{\odot}}\ h^{-1}$). {\sc enzo} is an AMR code, but this tool has not been
used in this project. This choice was made because our work is focused
mainly on mildly overdense gas, with a baryonic overdensity $\delta =
\rho / (\Omega_{\rmn b} \rho_{\rmn{cr}})$ in the range between 1 and
about 100 (see below in this Section). Here $\rho$ is the gas density and $\rho_{\rmn{cr}} = 3 H_0^2 /( 8
\pi G)\ (1 + z)^3$ is the critical density at redshift $z$. Properly
resolving this gas phase with AMR would be extremely volume-filling
and therefore very inefficient from a computational viewpoint, while
the adopted approach performs comparatively better, with an affordable
use of computational resources. The main drawback is the relatively
coarse resolution inside the haloes. With this
choice, the spatial resolution of our simulations is
$25\ \rmn{kpc}\ h^{-1}$. In a similar context, the same refinement
strategy was used in grid-based simulations of the WHIM by \citet{rhv07} and \citet{shs11}.

Four simulations have been compared in our study. They follow the
evolution of the same realisation of the initial conditions, but use
different sets of physical prescriptions. The first one is dubbed $NR$
for ``non-radiative'', in the sense that it does not include any
additional physics in solving the equations of fluid dynamics, whose
evolution is thus driven solely by gravity.

The second run is labelled with $F$ for {\sc fearless}\footnote{Fluid
  mEchanics with Adaptively Refined Large Eddy SimulationS.},
indicating with this acronym the use of turbulence subgrid scale (SGS)
model introduced by \citet{mis09} (see also
\citealt{snh06,ims10,isn11}). The turbulence SGS model is based on the
scale decomposition of the fluid dynamics equations in a resolved and
a small-scale part \citep{Germano1992}. A heuristic model is used to
predict the kinetic energy at the unresolved, subgrid scales (also
called SGS turbulence energy), and to couple it to the resolved
scales, through additional terms in the equations of fluid dynamics,
stemming from the filtering procedure. For the full set of equations
and other details we refer to \citet{mis09} and \citet{isn11}. One of
the main features of {\sc fearless}, namely the adjustment of the
energy budget to the changes in spatial resolution of the grid, is not
used in this work (with the exception of the runs presented in Section \ref{feedback}), because the computational grid is static and not
adaptive. We also address the reader to \citet{mis09} for numerical tests about the effect of resolution on the SGS turbulence, in simulations of forced isotropic turbulence in a periodic box.  

The third run is indicated by $HC$ (for ``Heating-Cooling''), because it implements the
equilibrium cooling model described in \citet{kwh96}, together with a
uniform UV background \citep{hm96}. The routines to implement the
effect of radiative cooling and UV background come
originally from the {\sc gadget-2} code \citep{springel05} and were
added into {\sc enzo} for the study of the Lyman-$\alpha$ forest
performed by \citet{rhv07}. The He heating rates have been multiplied
by a factor 3.3 in order to obtain an IGM temperature more in
agreement with observations \citep{v09}.  No feedback mechanism is
used in this work, but the cooling catastrophe at the centre of haloes
is avoided by imposing a pressure floor as in \citet{mba01}. Finally,
in our fourth run, labelled $FHC$, the turbulence SGS model, the UV
heating and the radiative cooling routines are used together.  The
summary of the runs performed is given in Table \ref{tab1sims}.

\begin{table}
\caption{Summary of the four simulations performed in the present
  work, showing the physical modules used in them. All runs follow the evolution of a cosmological volume with size of
  $10\ \rmn{Mpc}\ h^{-1}$ on a side with $400^3$ cells. The UV background
  used follows the implementation described in \citet{hm96}, the cooling is treated according to \citet{kwh96}, while for the turbulence SGS model we refer to \citet{mis09}.}
\centering
\begin{tabular}{ccc}
\hline
Simulation &  SGS turbulence model & Cooling and \\
 & & UV background \\ 
\hline 
\it{NR} & $ \times$ & $\times$
\\
\it{F} & \checkmark & $\times$  
\\
\it{HC} & $ \times$ &  \checkmark 
\\ 
\it{FHC} & \checkmark & \checkmark
\\ 
\label{tab1sims}
\end{tabular}
\end{table}

The analysis in this work will be focused mainly on gas phases with
selected features. Even without studying the process of metal
enrichment, one can infer a link to observations by probing the gas
with properties (density, temperature) which are more relevant for a
comparison with the observables.  The question at this point is: which
gas phases are traced by the available observables? Following the
arguments of \citet{rsb01}, it turns out the C {\sc iv} absorption
traces gas with overdensity from a few times the mean density of the
universe to about 200, namely a phase that is representative of the
gas surrounding galactic haloes. On the other hand, the O {\sc vi}
absorbers traces gas with a slightly lower overdensity ($\rho /
\overline{\rho} \lesssim 100$; \citealt{odk12,trs11}).  Without
attempting a detailed comparison of our results with the above
mentioned investigations, we will analyse mainly two gas phases:
\begin{itemize}
\item a diffuse component, which in the following will be simply referred to as IGM, with baryon overdensity $\delta$ between 1 and 50. We are well aware that, in literature, the term  IGM is used in a more generic way than in our present definition; 
\item a WHIM component, with temperature $T$ between $10^5$ and $10^7\ {\rm K}$. This component is representative of shock-heated gas, located closer to collapsed structures.
\end{itemize}
As we will see in the next Section, these two definitions do not divide the baryonic matter in an exclusive way: there is gas which can belong both to the IGM and the WHIM phase.

\section{Results}
\label{results}

\subsection{General features of the simulations}
\label{general}

\begin{figure*}
\centering
\includegraphics[width=0.9\linewidth,clip]{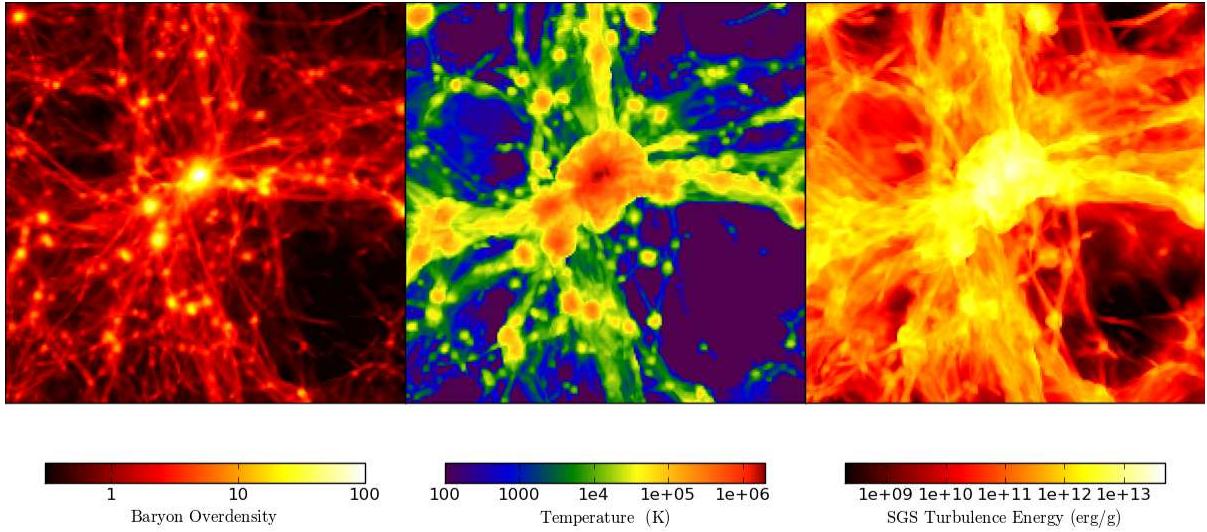}
\caption{The panels show a projection with size of $5\ \rmn{Mpc}\ h^{-1}$ on a side, centred on the most massive halo of the computational domain ($M_{200} = 1.33 \times 10^{12}\ {\rm M_{\odot}}\ h^{-1}$), at the final redshift $z = 2$, for the simulation $F$. From left to right, the panels show baryon overdensity, gas temperature and SGS turbulence energy. The quantities are colour-coded, according to the colour bars shown under the panels. The projections of density and temperature for the run $NR$ do not differ significantly from the ones shown here.}
\label{proj-f}
\end{figure*}

\begin{figure*}
\centering
\includegraphics[width=0.9\linewidth,clip]{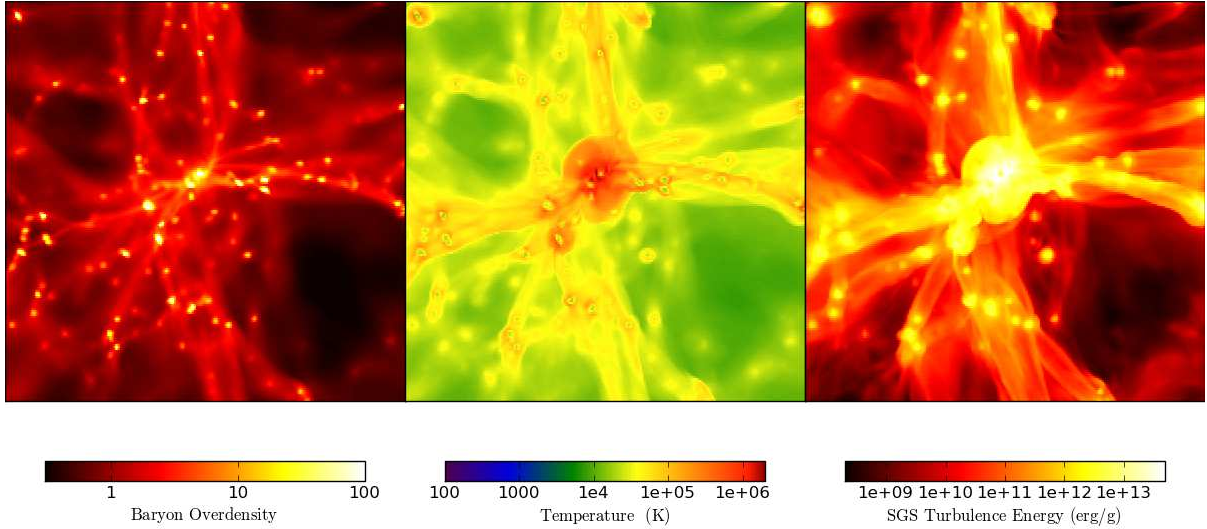}
\caption{Same of Fig.~\ref{proj-f}, for the run $FHC$. The projections of density and temperature for the run $HC$ do not show significant differences from the ones shown here.}
\label{proj-fhc}
\end{figure*}

\begin{figure*}
\centering
\includegraphics[width=0.6\linewidth,clip]{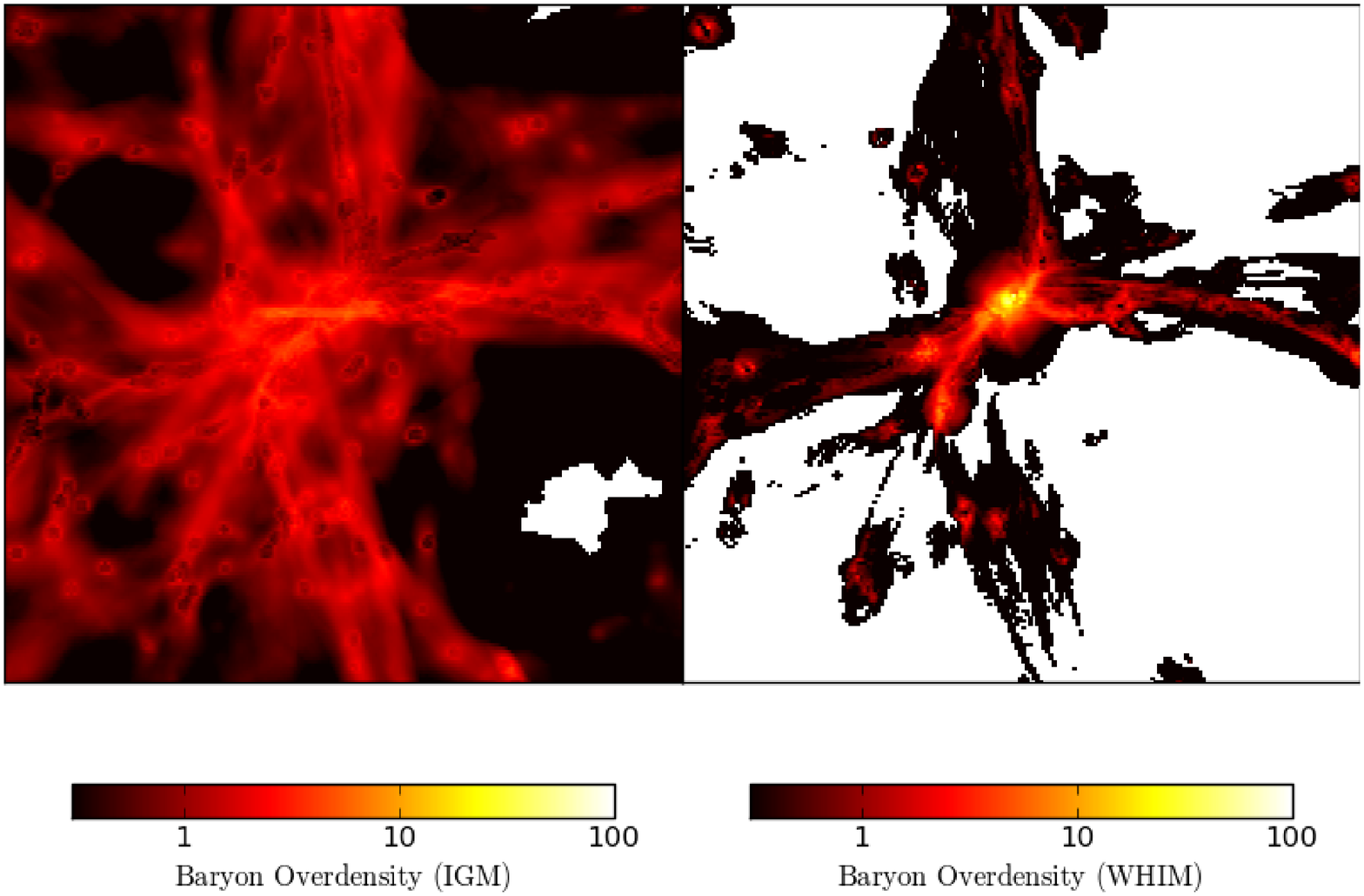}
\caption{Projections of baryon overdensity for the run $FHC$ at $z = 2$, like the left panel in Fig.~\ref{proj-fhc} but showing only gas belonging to the IGM (left) or to the WHIM phase (right panel) For sake of comparison, the colour bars have the same range as in Figs.~\ref{proj-f} and \ref{proj-fhc}.}
\label{proj-phases}
\end{figure*}

For a first contact with the simulation data, in Figs.~\ref{proj-f}
and \ref{proj-fhc} we present projections at $z=2$, centred on the most
massive halo evolved in the computational box, for the simulations $F$ and $FHC$. The runs $NR$ and $HC$ present only minor morphological differences with respect to the runs $F$ and $HC$, respectively, thus their projections are not shown here.

A prominent effect is the sizable role of the UV
heating and radiative cooling in shaping the gas evolution. Simulations
$HC$ and $FHC$ have a smaller number of
filamentary structures, when compared with the $NR$ or $F$ simulations. 
This is the consequence of the efficient cooling, which makes gas
condensing 
in galaxies, thus removing it from filaments. In a similar way
the denser structures corresponding to haloes are affected, with
cooling making them more compact and reaching higher densities. 

\begin{figure}
  \resizebox{\hsize}{!}{\includegraphics{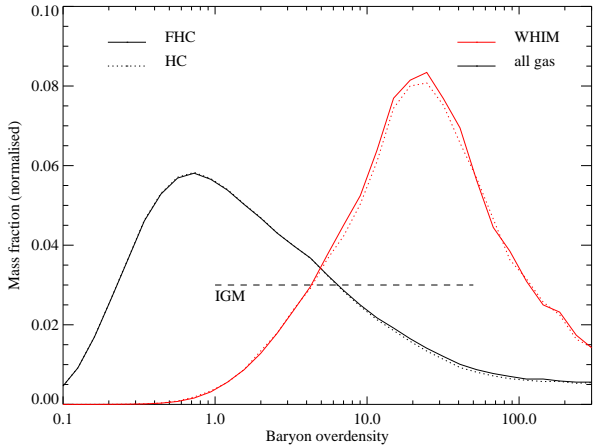}}
  \caption{Mass distribution function for all gas (black lines) and the WHIM phase (red lines) as a function of baryon overdensity. The solid lines indicate the $FHC$ run, and the dotted lines the $HC$ run, both at $z = 2$. The dashed horizontal line marks the overdensity range of the IGM phase.}
  \label{mass-pdf}
\end{figure}

In Fig.~\ref{proj-phases} we show the gas belonging to the IGM and
WHIM phase separately, for the run $FHC$. Clearly, the IGM gas is
associated to the filamentary structure and to the diffuse medium,
while the WHIM consists of gas surrounding more massive halos and
denser filaments. The plot shows also that the IGM and the WHIM are
not separated by their definitions, i.e.~there is IGM gas also
belonging to the WHIM phase. This can be seen even more easily in the
mass distribution function of the runs $HC$ and $FHC$, shown in
Fig.~\ref{mass-pdf}, for all gas (black lines) and the WHIM phase (red
lines). A significant fraction of the WHIM belongs to the IGM, but the
peak of its mass distribution is found at relatively large
overdensities ($\delta \sim 20$), consistently with what observed in
Fig.~\ref{proj-phases}.

\begin{figure}
  \resizebox{\hsize}{!}{\includegraphics{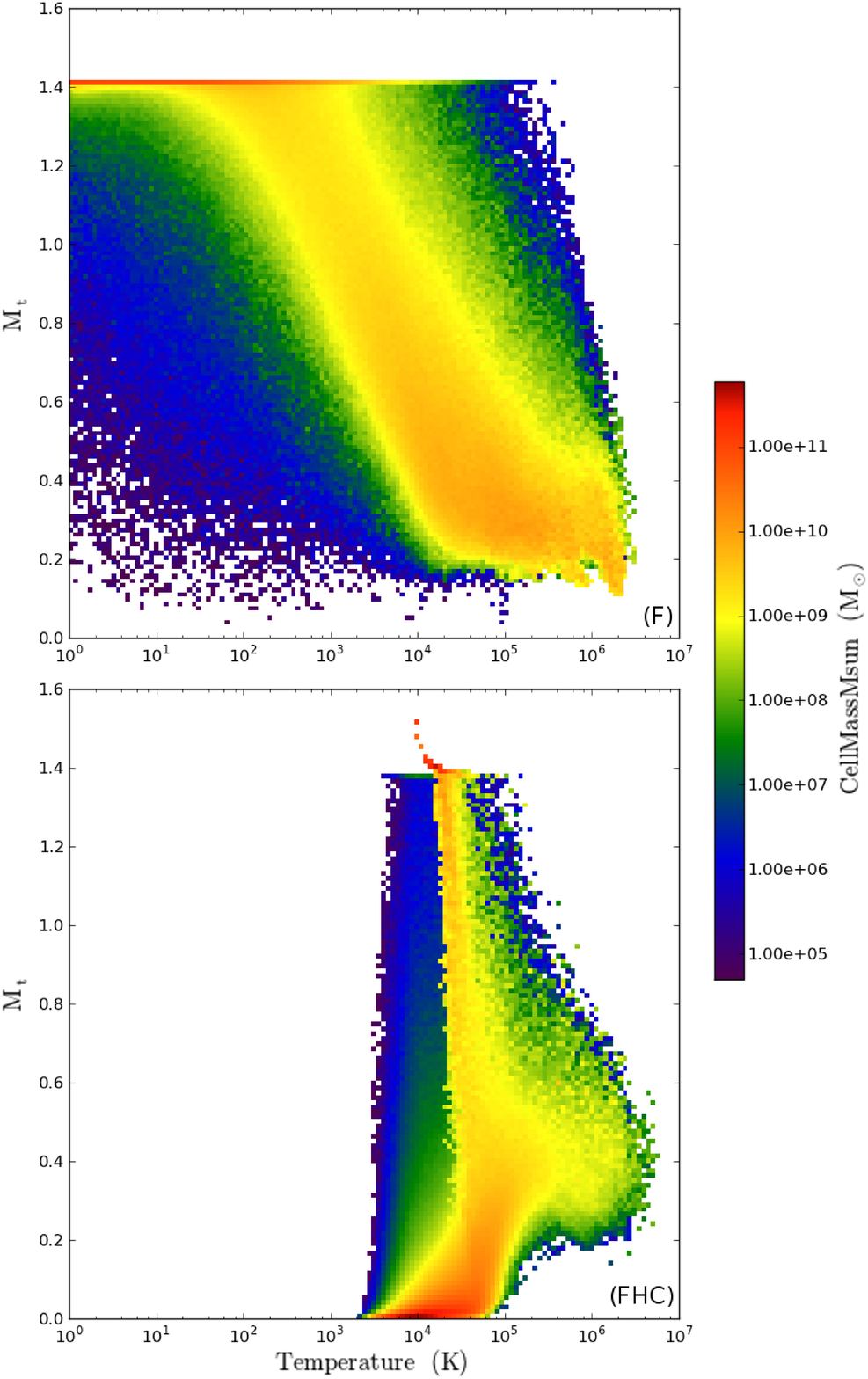}}
  \caption{Two-dimensional mass distribution function of the SGS turbulence Mach number $M_{\rm t}$ as a function of the gas temperature $T$, for the runs $F$ (upper panel) and $FHC$ (lower panel), at $z = 2$.}
  \label{mat-temp}
\end{figure}

It has been mentioned that the use of the SGS model does not have any significant impact on the global dynamics of the structure formation. The reason for this can be understood from
the Mach number of the SGS turbulence, which is defined as
\begin{equation}
\label{mturb}
M_{\rm t} = \frac{q}{c_{\rm s}} \,\, ,
\end{equation}
where $q = (2e_{\rm t})^{1/2}$ is the SGS turbulence velocity, $e_{\rm
  t}$ is the SGS turbulence energy, and $c_{\rm s}$ is the sound
speed. As we will see in more detail in Section \ref{sgs-properties},
on average value of $M_{\rm t}$ is small (between 0.02 and 0.3) for the gas in
the overdensity range $1 < \delta < 50$. Interestingly, the average
$M_{\rm t}$ is larger for the WHIM ($\langle M_{\rm t} \rangle =
0.43$; cf.~Fig.~\ref{mat-temp}, lower panel), so we expect that
turbulence (and its modelling) are more relevant for this gas
phase. We will come back in the next Sections on this key point of our
analysis.

\begin{figure*}
\centering
\includegraphics[width=0.9\linewidth,clip]{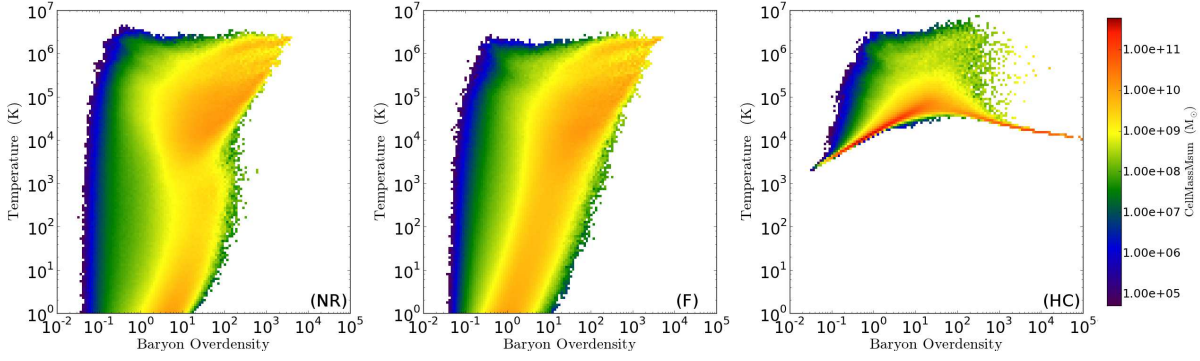}
\caption{The three panels show the two-dimensional mass distribution function of gas temperature as a function of baryon overdensity, at $z = 2$. The simulations are indicated by the labels in the lower right corner of each panel. The panel for run $FHC$ is not shown, because it is basically identical to that for run $HC$.}
\label{t-rho-panels}
\end{figure*}

A more quantitative comparison of the performed simulations is provided by
the two-dimensional mass distribution function of temperature as a
function of baryon overdensity, in Fig.~\ref{t-rho-panels}. 

Here a difference between run $NR$ and $F$ can be seen, because the
latter contains more mass at $T \la 10^4\ \rm{K}$. This is a side
effect of the SGS model, that heats unphysically some pristine cold
gas. From equation (\ref{mturb}) it follows that in the cold gas,
which has very low sound speed, even a moderate SGS turbulence may
result in a large $M_{\rm t}$. This is indeed the case for the gas at
low temperature in the $F$ run, as it can be better seen from the two
dimensional mass distribution of $M_{\rm t}$ as a function of $T$ in
Fig.~\ref{mat-temp}. In the upper panel, a substantial amount of gas
at low temperature has $M_{\rm t} = 2^{0.5}$, which is the threshold
value allowed by the safeguarding mechanism implemented in the SGS
model \citep{mis09}, which does not provide reliable results for
larger values of $M_{\rm t}$. The UV heating included in the run $FHC$
completely solves the problem, by removing the unshocked gas from the
cold phase. The SGS turbulent Mach number still reaches very large
values but only in the densest part of the halo cores, and we are not
concerned with the analysis of those under-resolved regions.

\begin{figure}
  \resizebox{\hsize}{!}{\includegraphics{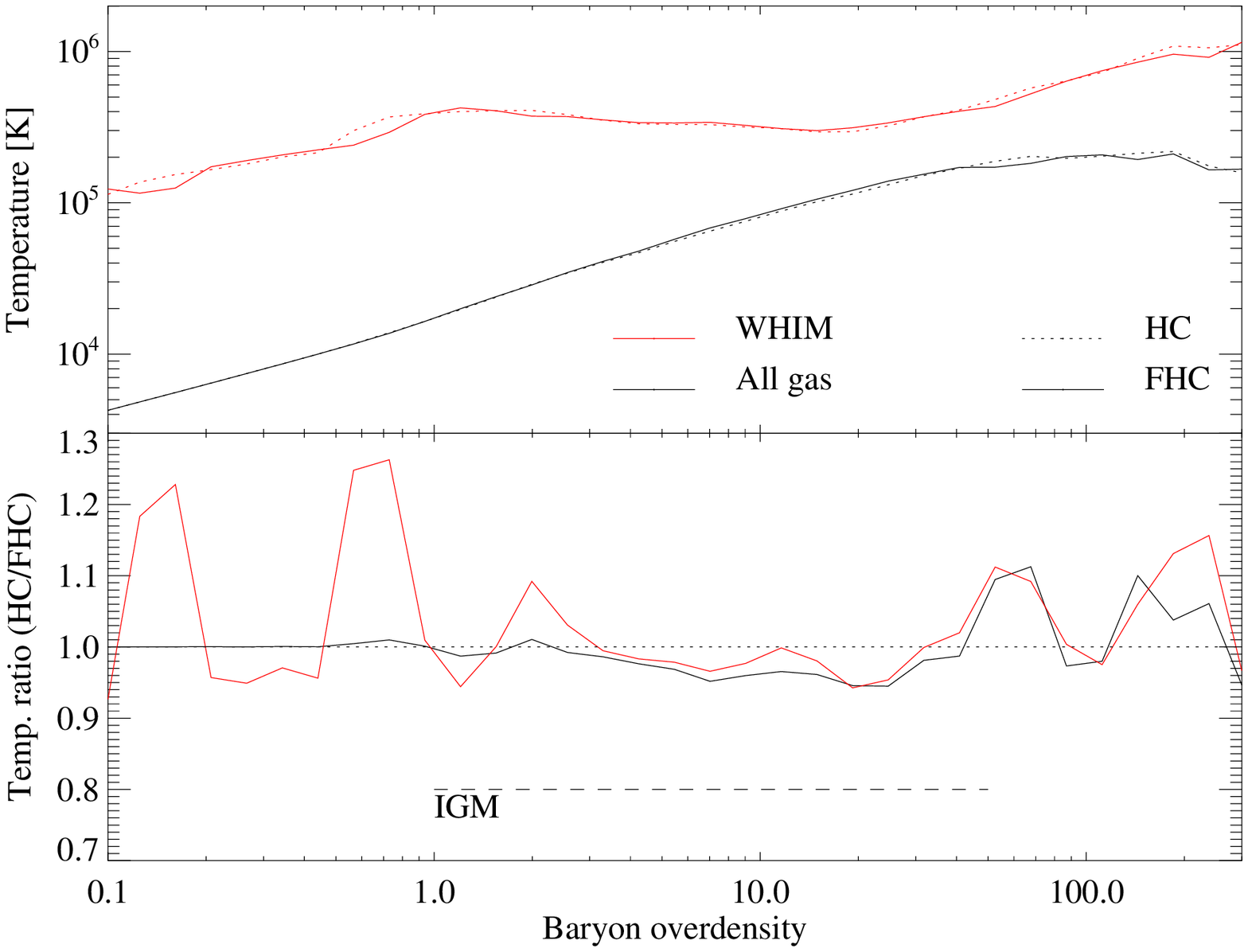}}
  \caption{Upper panel: mass-weighted average of the gas temperature as a function of the baryon overdensity $\delta$ at $z = 2$, for the runs $FHC$ (solid lines) and $HC$ (dotted lines), for all gas (black lines) and the WHIM phase (red lines). Lower panel: ratio between the $T$ average of the run $HC$ and $FHC$ for all gas (black line) and the WHIM (red line). The dashed horizontal line indicates the IGM overdensity range.}
  \label{t-rho-1d}
\end{figure}

The two-dimensional $T$--$\delta$ mass distribution function of the
run $FHC$ is not included in Fig.~\ref{t-rho-panels}, because the differences with the run $HC$ cannot be easily appreciated with this kind of diagnostic. 
A better visualisation is provided by the average temperature as a function of $\delta$, shown in Fig.~\ref{t-rho-1d}.

We observe that the average temperatures agree within 10 per cent,
with the exception of the two fluctuations for the WHIM at $\delta <
1$, which have a negligible impact because of the very low mass of the
WHIM gas at that density (cf.~Fig.~\ref{mass-pdf}).  In the
overdensity range $5 < \delta < 40$ the average temperature for the
run $FHC$ is slightly larger, especially for the IGM but also for the
WHIM. A similar increase in temperature, due to the effect of
including subgrid turbulence, has been already observed in the
simulations of \citet{mis09}, and is caused by the turbulent
dissipation, which convert a part of the unresolved kinetic energy to
internal energy.

The peak of the WHIM mass distribution lies in this overdensity range. This is probably related to the fact that, in Fig.~\ref{mass-pdf}, the WHIM mass fraction around the peak is slightly larger in the $FHC$ run than in the $HC$. For this reason, it is not surprising that the SGS model is especially relevant for the energy budget of gas in that phase.

At overdensities between $\delta = 40$ and 100 the behaviour is
opposite, and the run $HC$ has an average temperature up to $10$ per
cent larger. In Section \ref{haloes} this will be interpreted in terms
of the buffer effect introduced by the turbulence SGS model, and
examples of both this temperature trend and the previous one are
discussed.

Finally, the properties of the average temperature at even larger overdensities are more complicated. Most of the gas with $\delta \gtrsim 100$ belongs to haloes, whose properties are not addressed in this study. An analysis will be performed locally in the vicinity of a few selected haloes, in Sections \ref{haloes} and \ref{los}. Further properties of the SGS turbulence energy are
presented in the next Section.

\subsection{Properties of subgrid turbulence}
\label{sgs-properties}

We describe here the main features of the SGS turbulence, as inferred from run $FHC$. In this computational setup, by subgrid scales we mean length scales just smaller than the spatial resolution ($25\ \rmn{kpc}\ h^{-1}$). If not otherwise specified, the analysis will not be extended to run $F$, because $FHC$ includes the additional physics modules (cooling, UV heating) which are thought to be important for a plausible description of the gas physics at these scales.  

\begin{figure}
  \resizebox{\hsize}{!}{\includegraphics{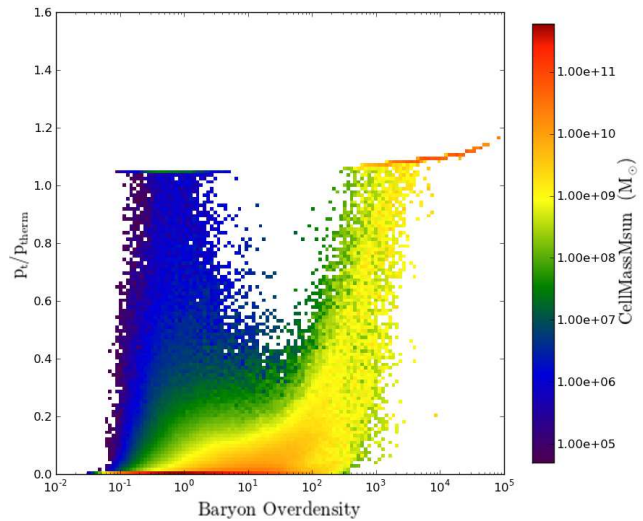}}
  \caption{Two-dimensional mass distribution function of the pressure ratio $p_{\rm t} / p_{\rm therm}$ as a function of the baryon overdensity $\delta$, for the run $FHC$, at $z = 2$.}
  \label{pratio-2d}
\end{figure}

In Section \ref{general} the turbulence Mach number $M_{\rm t}$ was introduced for a first, brief introductory discussion of unresolved turbulence. In the following we define a further turbulence diagnostic, the ratio between the SGS turbulent pressure and thermal one $p_{\rm t} / p_{\rm therm}$. Defining the SGS turbulent pressure as
\begin{equation}
\label{pturb}
p_{\rm t} = \frac{2}{3} \rho e_{\rm t} \,\, ,
\end{equation}
the ratio is then given by
\begin{equation}
\label{pratio}
p_{\rm t} / p_{\rm therm} = \frac{\frac{2}{3} \rho e_{\rm t}}{\rho e_{\rm int} (\gamma - 1)} \,\, ,
\end{equation}
which, for a value of the ratio of specific heats $\gamma = 5/3$ in the equation of state used at the denominator, reduces to the ratio between the SGS
turbulent energy and the internal energy. Moreover, from a comparison of
equations (\ref{mturb}) and (\ref{pratio}), it follows:
\begin{equation}
\label{pratio-rel}
p_{\rm t} / p_{\rm therm} =  \frac{\gamma (\gamma -1)}{2} M^2_{\rm t}
\end{equation}

The two-dimensional mass distribution function of the pressure (or
energy) ratio as a function of the baryon overdensity is shown in
Fig.~\ref{pratio-2d}. The value of the pressure ratio is generally low
at small $\delta$ and grows for $\delta \gtrsim 10$. Additionally, two
prominent features can be seen both at low and high $\delta$. The
first one is related to the problem of the unphysical SGS turbulence
production in low-density gas, discussed in Section
\ref{general}. Here, again, the flattening of the peak is caused by
the safeguarding mechanism of the SGS model \citep{mis09}. We notice
that the mass of the involved gas is safely negligible, in comparison
with run $F$ (Fig.~\ref{mat-temp}, upper panel). The flat peak at
higher density is related to the gas at the centre of halo cooling
cores, and in particular to the pressure floor introduced there. This
feature does not concern our analysis, because the innermost regions
of the haloes will not be studied in this work.

\begin{figure}
  \resizebox{\hsize}{!}{\includegraphics{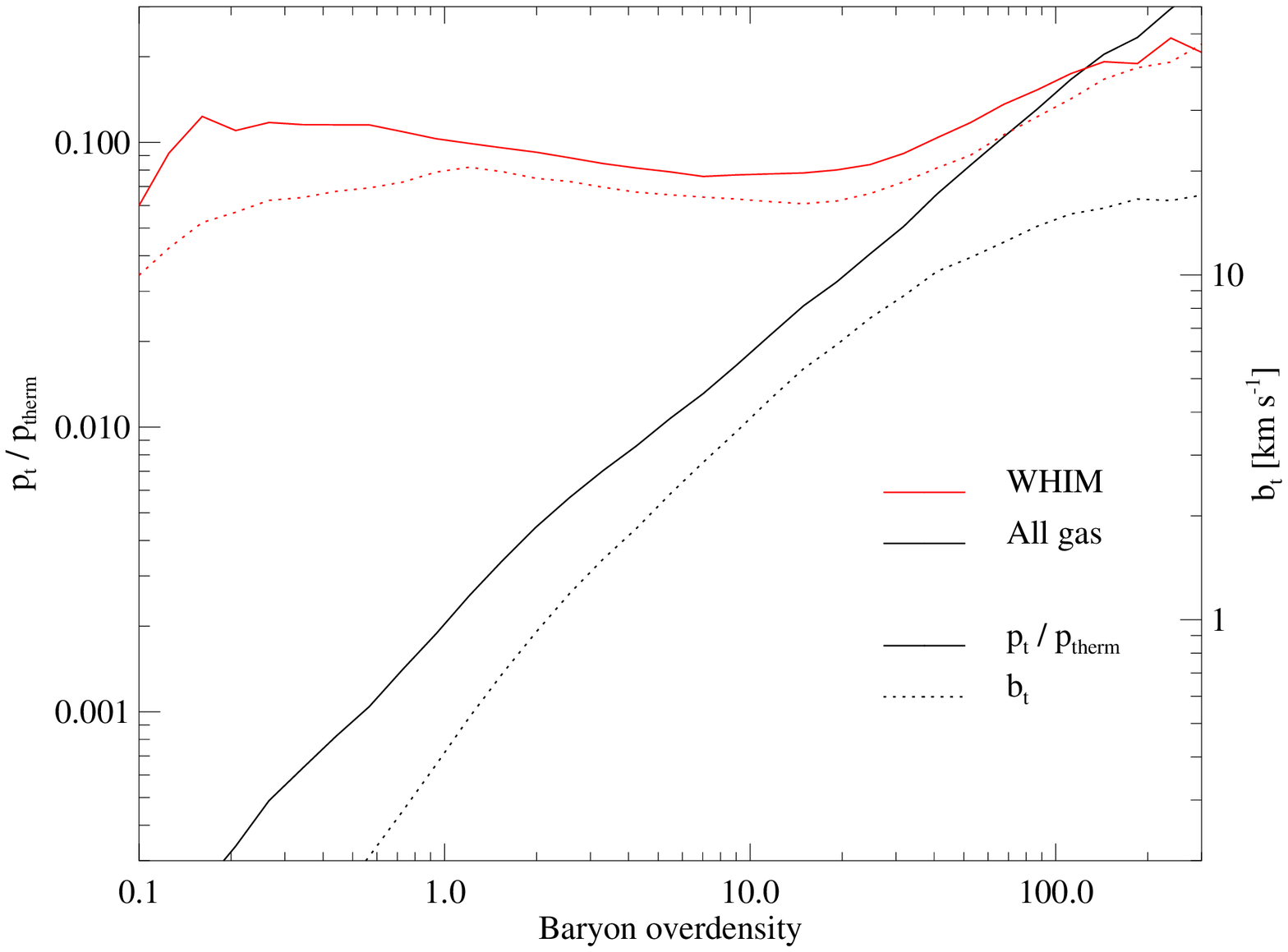}}
  \caption{Average of the pressure ratio $p_{\rm t} / p_{\rm therm}$ (solid lines, $y$-axis on the left-hand side) and of turbulent Doppler broadening $b_{\rm t}$ (dotted lines, $y$-axis on the right-hand side) for all gas (black lines) and the WHIM gas (red lines).}
  \label{pratio-1d}
\end{figure}

The average values of the pressure ratio as a function of baryon overdensity are plotted in Fig.~\ref{pratio-1d}. For the IGM the mass-weighted average of $p_{\rm t} / p_{\rm therm}$ grows from 0.002 to 0.08. At the upper edge of the overdensity range, the SGS turbulent pressure starts being globally sizeable and, locally, there are regions where $p_{\rm t} / p_{\rm therm}$ can reach the value of 0.4 (Fig.~\ref{pratio-2d}). The WHIM phase has a markedly different and almost flat trend with overdensity, with typical values of $p_{\rm t} / p_{\rm therm}$ around 0.1. As already anticipated in Section \ref{general}, turbulence is generally important in the WHIM.
The reason for this is that low-density WHIM is contributed by gas
that is shock heated to warm-hot temperatures in the outer regions of
filaments and haloes, as a consequence of its diffuse accretion onto
such structures. This is the regime where turbulence is
injected most efficiently, thus making its impact on the WHIM more
pronounced than in the gas at comparable density, but lower temperature. 

In Fig.~\ref{pratio-1d} a further diagnostic of turbulent gas motions is introduced for its relevance in the comparison with observations, namely the turbulent Doppler parameter $b_{\rm t}$, derived from the SGS pressure like in \citet{ef11}:
\begin{equation}
\label{bturb}
b_{\rm t} =  \sqrt{\frac{p_{\rm t}}{\rho}} = \frac{q}{\sqrt{3}}
\end{equation}
In our case, $b_{\rm t}$ probes length scales at or just below the
spatial resolution of the simulations ($25\ {\rm kpc}\ h^{-1}$). The
turbulent Doppler parameter $b_{\rm t}$ for the IGM grows from 0.5 to
$10\ {\rm km\ s^{-1}}$, with a volume-weighted average value of $1.02\
{\rm km\ s^{-1}}$, while for the WHIM the average is $18.2\ {\rm km\
  s^{-1}}$.  

Although at high overdensity the same average values are
reached in Fig.~\ref{pratio-1d} for IGM and WHIM, the different mass
distributions in Fig.~\ref{mass-pdf} make the large values in the IGM
not relevant on average for this phase.

The values of $b_{\rm t}$ seen above have to be put in the proper context, by comparing them with the thermal broadening $b_{\rm therm} = (2 k_{\rm B} T / m_{\rm H})^{1/2}$, where $k_{\rm B}$ is the Boltzmann constant and $m_{\rm H}$ is the mass of hydrogen, the atom that we are considering for simplicity. In this case, the Doppler broadening ratio can be expressed as:
\begin{equation}
\label{dratio}
\frac{b_{\rm t}}{b_{\rm therm}} = \sqrt{\frac{p_{\rm t} m_{\rm H}}{2 \rho k_{\rm B} T}} =  \sqrt{\frac{1}{2\mu}} \sqrt{\frac{p_{\rm t}}{p_{\rm therm}}} \simeq 0.91 \sqrt{\frac{p_{\rm t}}{p_{\rm therm}}}
\end{equation}
where we made use of the expression of the thermal pressure for the ideal gas $p_{\rm therm} = \rho T k_{\rm B} / (\mu m_{\rm p})$, with the approximation for the proton mass $m_{\rm p} \sim m_{\rm H}$; $\mu = 0.6$ is the mean molecular weight in a.m.u..

\begin{figure}
  \resizebox{\hsize}{!}{\includegraphics{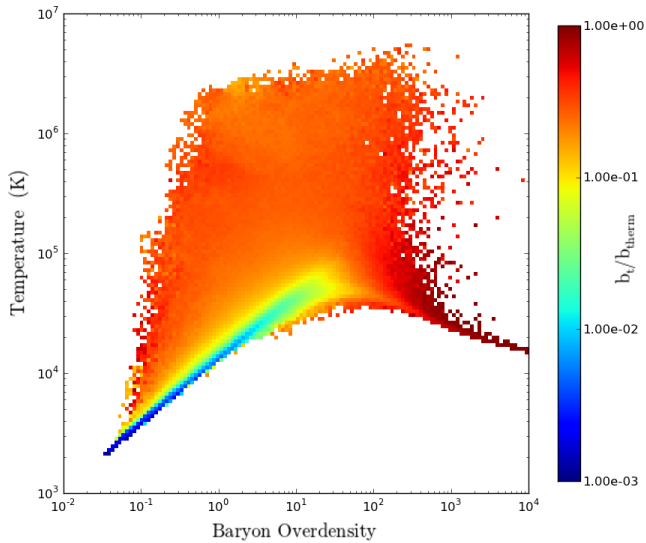}}
  \caption{Two-dimensional distribution function of the average of the Doppler broadening ratio $b_{\rm t}/b_{\rm therm}$ in the $T-\delta$ plane.}
  \label{bratio-2d}
\end{figure}

\begin{figure}
  \resizebox{\hsize}{!}{\includegraphics{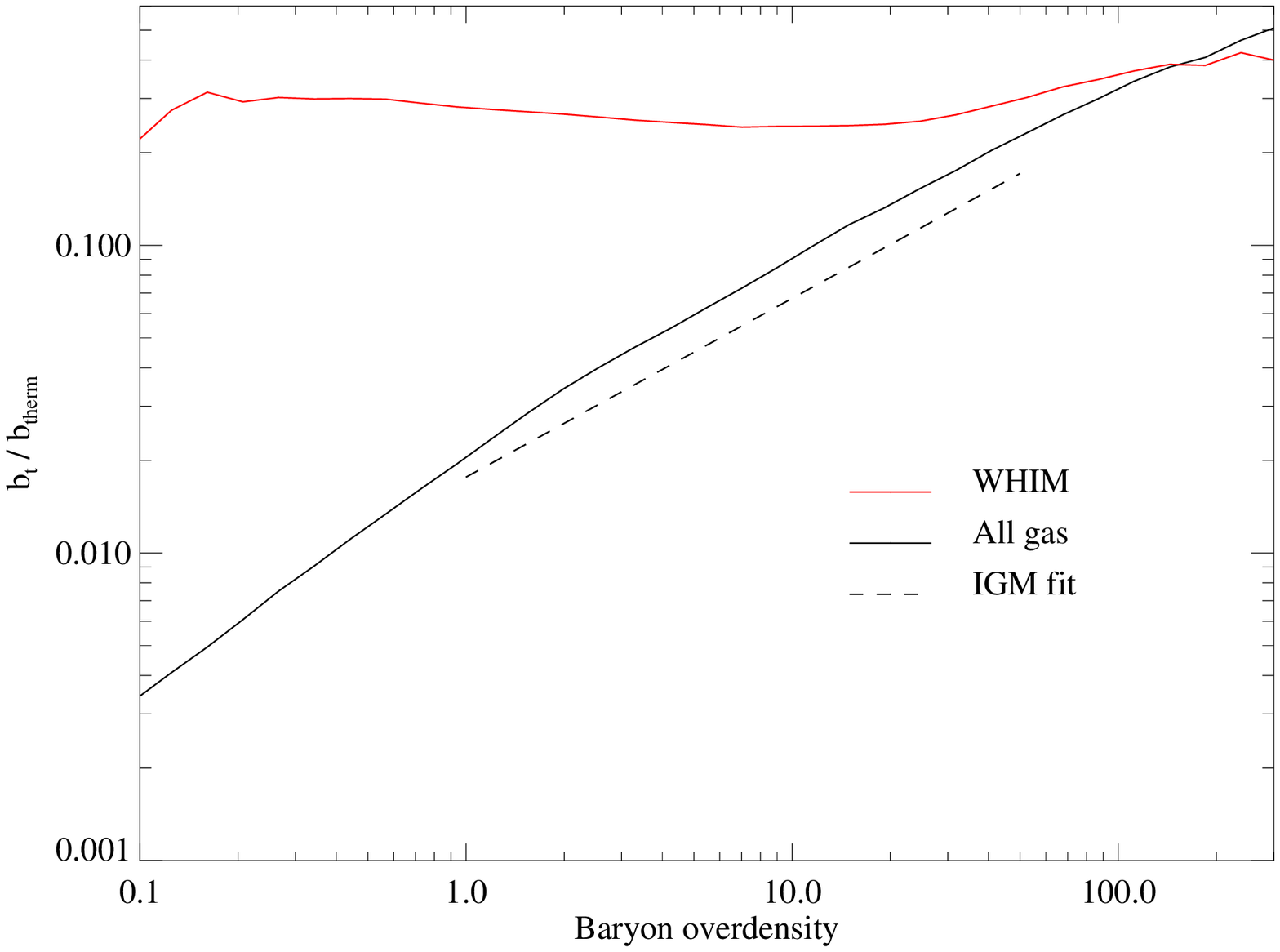}}
  \caption{Averages of the Doppler broadening ratio $b_{\rm t}/b_{\rm therm}$ for all gas (black solid line) and the WHIM gas (red line) as a function of the baryon overdensity. The dashed line shows the fit in the IGM overdensity range $b_{\rm t}/b_{\rm therm} = 0.023 \times \delta^{0.58}$, slightly shifted downwards for ease of visualisation.}
  \label{bratio-fit}
\end{figure}

From the two-dimensional distribution function of $b_{\rm t}/b_{\rm therm}$ (Fig.~\ref{bratio-2d}) it is evident that the Doppler broadening ratio is much larger in the shock-heated WHIM gas than in the low-density photoionised phase. This is further confirmed by the averages shown in Fig.~\ref{bratio-fit}. Given the strong link emphasised in equation (\ref{dratio}), the overall trends for WHIM and IGM are similar to the pressure ratio in Fig.~\ref{pratio-1d}. The volume-weighted averages of $b_{\rm t}/b_{\rm therm}$ are 0.034 and 0.26 for the IGM and WHIM phase, respectively. In Fig.~\ref{bratio-fit} we show that the increase of $b_{\rm t}/b_{\rm therm}$ with $\delta$ is fitted, in the IGM overdensity range, by the function $b_{\rm t}/b_{\rm therm} = 0.023 \times \delta^{0.58}$.

The analysis in the vicinity of massive haloes, which will be presented in the next Sections, concerns mostly WHIM gas, for which the effect of modelling unresolved turbulence by a SGS model are the most relevant.

\subsection{Properties of turbulence in the vicinity of selected haloes}
\label{haloes}

We present here a qualitative comparison of the thermal and turbulent properties of the gas around the second most massive halo formed in our computational box (chosen for exemplifying reasons), with mass of $1.06 \times 10^{12}\ {\mathrm M_{\odot}}\ h^{-1}$ at $z = 2$, for the runs $HC$ and $FHC$. This analysis provides a schematic picture of the most important effects of the SGS model on the gas physics in the outskirts of galaxy-sized haloes and, in prospect, of turbulence driving in this framework. 

\begin{figure*}
\centering
\includegraphics[width=0.6\linewidth,clip]{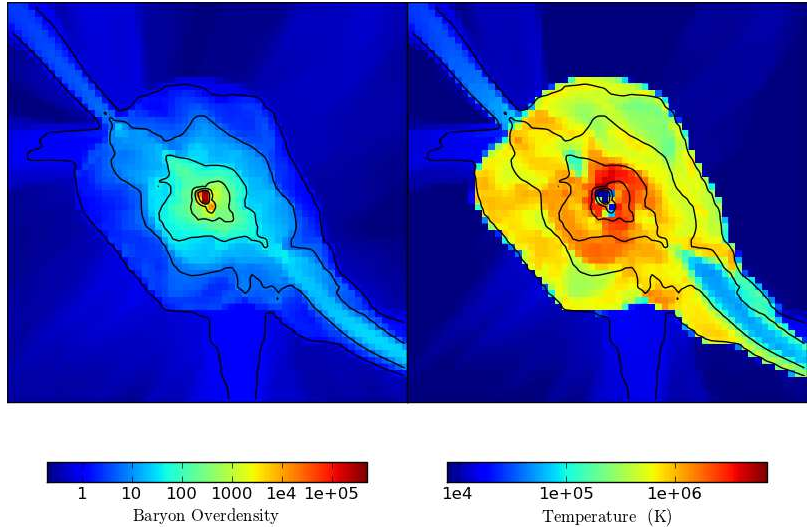}
\caption{Slices normal the $x$-axis, with size of $1.5\ {\rm Mpc}\ h^{-1}$ on a side, centred on a massive halo at $z = 2$, for the run $HC$. The panels refer to baryon overdensity and temperature, from left to right, and are colour-coded according to the colour bars on the bottom. Black density contours are overplotted on both panels.}
\label{second-hc}
\end{figure*}

\begin{figure*}
\centering
\includegraphics[width=0.9\linewidth,clip]{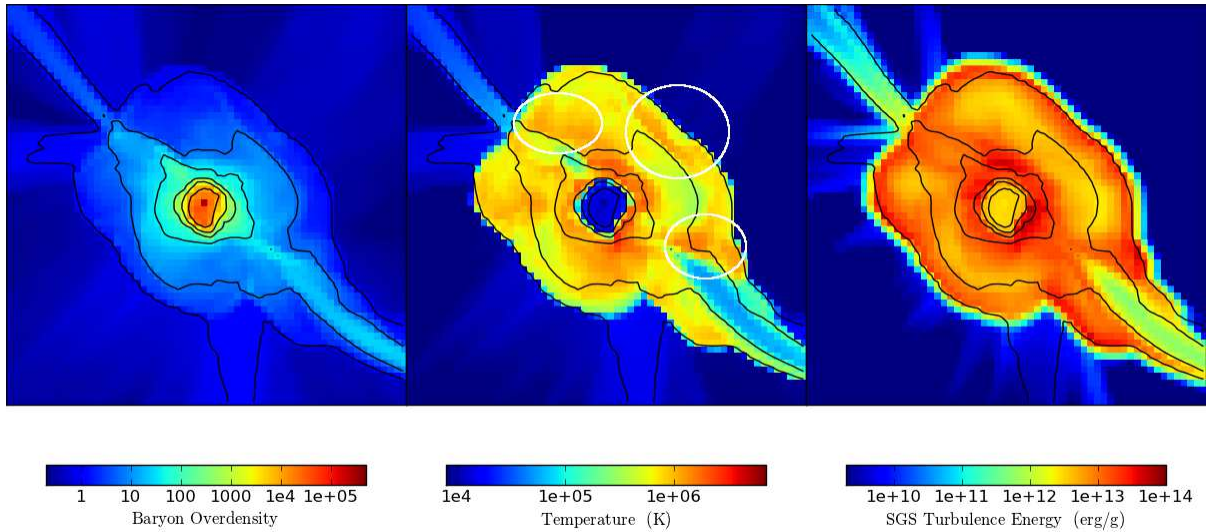}
\caption{Same as Fig.~\ref{second-hc}, but for run $FHC$. A panel for the SGS turbulence energy is added at the right-hand side. The regions enclosed in the white ellipses have a temperature larger than in the run $HC$, as discussed in the text.}
\label{second-fhc}
\end{figure*}

Figs.~\ref{second-hc} and \ref{second-fhc} show a series of slices of gas density, temperature and (for the run $FHC$) SGS turbulence energy. The white ellipses in Fig.~\ref{second-fhc} put into evidence regions where, in comparison with the run $HC$, the temperature is higher. In two cases, the temperature increase occurs in regions where an inflowing filament (from the upper left to the lower right corner in the slices) drives shear flows in the surrounding WHIM gas. This process produces turbulence, as observed in simulations at cluster scales by \citet{pim11}. For the central ellipse, a visual inspection suggests that turbulence is injected in that region mostly by baroclinic and compressive effects at the curved external shock (e.g., \citealt{ib12}). In all these cases, modelling the unresolved turbulence brings to a higher temperature in the region around the inflowing filament or in the post-shock region. In \citet{mis09}, it is argued that the increase in temperature is caused by the additional dissipation of SGS energy.

On the other hand, the use of the turbulence SGS model can also result in a temperature \emph{decrease} of the gas. In Fig.~\ref{second-fhc}, the most apparent case is in the surroundings of the halo centre. This example illustrates the role of the SGS model as an energy buffer inserted between the resolved kinetic and the internal energy component (see \citealt{ims10} for a graphical sketch of this concept). According to this interpretation, part of resolved kinetic energy is retained as unresolved SGS energy and is not directly dissipated to internal energy, resulting in some locations with colder gas in the simulation $FHC$. 

To sum up, a simple explanation would proceed as follows: it is conceivable that soon after turbulence injection, the inclusion of the turbulence SGS model would cause a reduction of temperature. On the other hand, subsequent dissipation of SGS turbulence could cause instead an increase of temperature with respect to the case in which no subgrid turbulence is included.
It would be interesting to understand under which conditions one of these two effects prevails, and to find a clear correlation between the variations of temperature in the runs $HC$ and $FHC$ and some quantity related to the SGS turbulence, such as the SGS energy, or any source term in its governing equation \citep{isn11}. 
Several tests (not included here) have not shown any correlation of this kind; this result is not surprising, given the highly nonlinear nature of the structure formation problem. For this reason, we will not try to bring this analysis beyond the mere qualitative level proposed in this Section. Further work, based on idealised simulation setups which are of simpler interpretation with respect to full cosmological simulations, might be able to shed light on this point.

\subsubsection{Effect of thermal and chemical feedback}
\label{feedback}

As already discussed in the Introduction, the simulations presented in this work do not include the effect of star formation and feedback. In this Section we will briefly report some results of runs in which modules for star formation, thermal feedback and metal enrichment are used in the simulation code. These results are intended to be an intermediate step towards a motivated implementation of kinetic feedback (at resolved and SGS scales) from galactic outflows, which is left for future work.

Different from the runs presented elsewhere in this paper, for the simulations shown here we used the {\sc enzo} code in its version 2.1, plus modifications for the implementation of the {\sc fearless} tool. The cosmological parameters, box size and initial/final redshifts are the same as in Section \ref{tools}. For a better resolution around galaxy-sized halos, the root grid is resolved with $128^3$ cells and the same number of N-body particles, and AMR up to three additional levels was used (effective spatial resolution of $9.7\ \rmn{kpc}\ h^{-1}$), with refinement criteria based on overdensity (cf.~\citealt{in08}) and refinement thresholds of 8 and 4 for DM and baryonic matter, respectively. These choices result in a spatial resolution for the WHIM gas that is comparable or better than in the runs presented in Section \ref{tools}; for the IGM gas, the same is true for $\delta \gtrsim 10$. Like in the previous runs, for the UV background the prescription of \citet{hm96} was followed.

Star formation and stellar thermal feedback (internal energy returned to the baryonic gas) was modelled according to the approach and the standard parameters used by \citet{shs11}. As for the metal feedback, two different strategies have been tested:
\begin{itemize}
\item together with thermal feedback, metal enrichment to the CGM is also considered, with the same scheme, efficiencies and yields as in \citet{shs11}. In this case (which will be dubbed as run $ThermMetal$), the radiative cooling is provided by a primordial chemistry network \citep{aaz97,aza97} coupled to the metal cooling prescriptions of \citet{gj07};
\item thermal feedback is used but, for a more consistent comparison with the other runs of this work, metal feedback is neglected. In this case, radiative cooling is computed using the analytical approximation from \citet{sw87}, assuming a fully ionised gas with half the solar metallicity. This run will be indicated as $Therm$.
\end{itemize}
 
\begin{figure*}
\centering
\includegraphics[width=0.9\linewidth,clip]{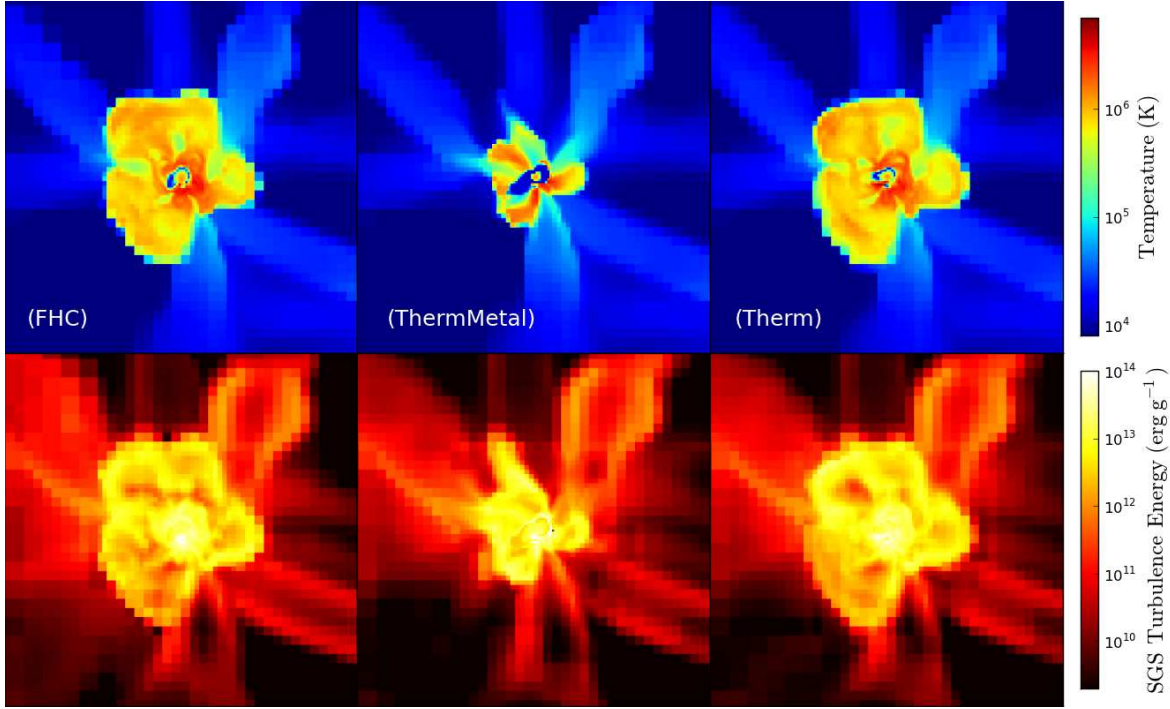}
\caption{Slices normal to the $x$-axis, with size of of $1.5\ {\rm Mpc}\ h^{-1}$ on a side, centred on a halo ($M = 8.02 \times 10^{11}\ {\mathrm M_{\odot}}\ h^{-1}$) at $z = 2$, for the three simulations described in Section \ref{feedback}. The slices show gas temperature (panels in the upper row) and SGS turbulence energy (lower row), colour-coded as in the colour bars on the right-hand side. The columns are ordered according to the different simulations, as indicated by the labels in the upper panels.}
\label{feedback-slices}
\end{figure*}

\begin{figure}
  \resizebox{\hsize}{!}{\includegraphics{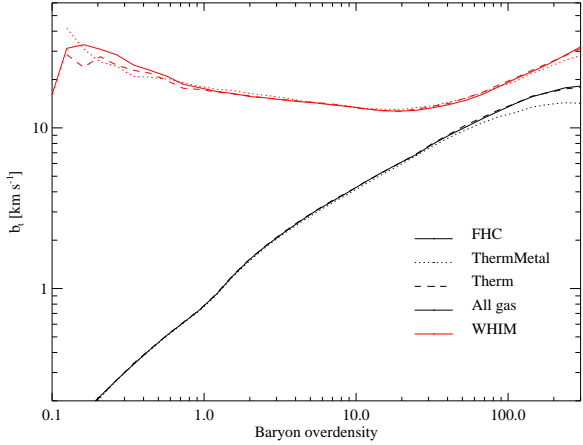}}
  \caption{Averages of the turbulent Doppler broadening $b_{\rm t}$ as a function as the baryon overdensity $\delta$, for the simulations described in Section \ref{feedback} (indicated by the legend). The group of black lines is for all gas, the red lines for the WHIM.}
  \label{1d-bratio-feedback}
\end{figure}

In Fig.~\ref{feedback-slices} these two feedback runs are compared with a simulation, labelled as $FHC$, in which only UV heating, radiative cooling \citep{sw87} and the turbulence SGS model are employed. The main conclusion that one can draw from the cross-comparison is that the thermal feedback alone has a limited impact on the CGM. Indeed, an analysis like the one performed in Fig.~\ref{pratio-1d} does not show any significant difference in the average of $b_{\rm t}$ as a function of $\delta$, between the runs $FHC$ and $Therm$ (Fig.~\ref{1d-bratio-feedback}).

Already a simple visual inspection of the central panels in Fig.~\ref{feedback-slices} suggests that the effect of metal feedback is more sizable. In particular, in run $ThermMetal$ metal cooling has a relevant impact on the morphology of the external shock, which moves much closer to the halo centre. This effect is due to the enhanced cooling rate contributed from metal line cooling. Since metals are not efficiently spread in the CGM by thermal feedback, this enhanced cooling mainly affects gas in the more central regions, where metals are in fact confined.

We decided not to include metal injection in our standard runs, for the following two reasons: firstly, the treatment of metals is not complete without a model for galactic winds, and such a kinetic feedback model is currently not implemented in the {\sc enzo} code. Secondly, we see in Fig.~\ref{1d-bratio-feedback} that, as a consequence of this lack of modelling of metal spread, the average of  $b_{\rm t}$ as a function of $\delta$ does not change much between the runs $FHC$ and $ThermMetal$, despite of the morphological differences. We leave the potential role of metals in this problem as potentially interesting, and worth to be taken into consideration for future work.

\subsection{Analysis of physical properties around massive haloes}
\label{los}

We now look into physical quantities related to volumes of
$1\ (\rmn{Mpc}\ h^{-1})^3$ centred around each of the three most massive
haloes, for all simulations. The following analysis is based entirely on the volume
centred on the most massive halo formed in the computational box ($M_{200} = 1.33 \times 10^{12}\ {\rm M_{\odot}}\ h^{-1}$ at $z = 2$), results for other haloes being very similar. Rather than analysing physical
  quantities related to subgrid scales, as done in the previous Sections, we
  extract the gas temperature, density and peculiar velocity fields
  from the simulated grid
    (i.e.~we do not interpolate physical values at grid
  points, but use the quantities as already provided by the simulation
  grid).  In this way we compare more quantitatively the physical
  state of the flow around these haloes in a relatively large
  cosmological volume. The analysis of the resolved velocity, in
  particular, complements the study of the SGS
  turbulence, and can support the results obtained in Section
  \ref{sgs-properties}.

\begin{figure*}
  \resizebox{\hsize}{!}{\includegraphics{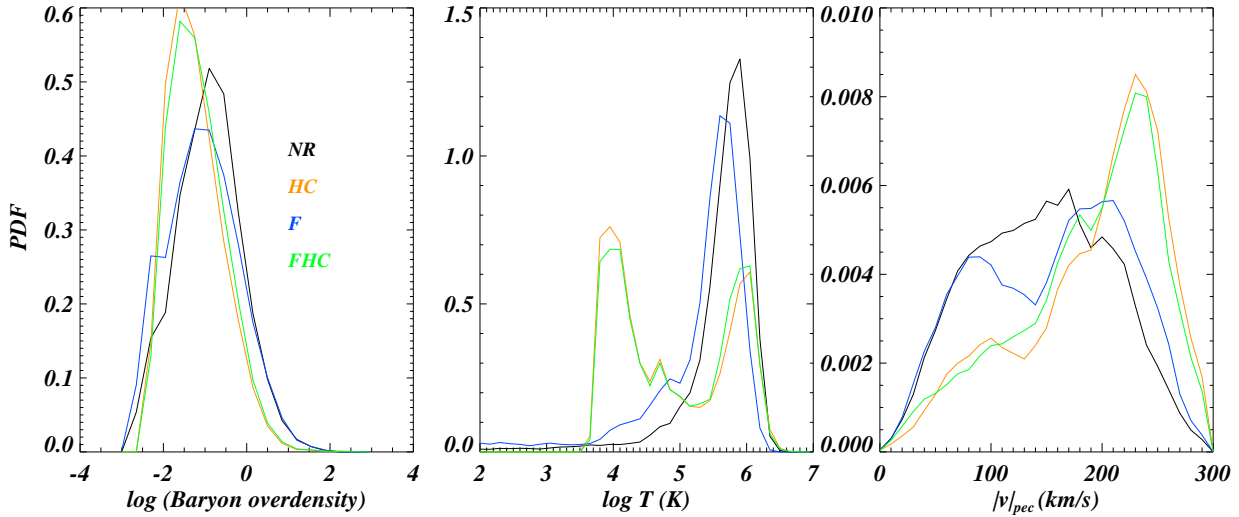}}
  \caption{Probability distribution function of the gas density,
    temperature and modulus of peculiar velocity field around the most massive halo
    for the four different runs investigated in this project:
    {\it NR} -- non-radiative model, {\it HC} -- run with UV
    background and cooling, {\it F} -- SGS turbulence model
    and {\it FHC} -- run with the UV background
    and turbulence SGS model (black, orange, blue and green curves, respectively).}
  \label{halo1}
\end{figure*}

In Fig.~\ref{halo1} we show the probability distribution functions of the gas
density (logarithmic units for the baryon  overdensity),
temperature and modulus of the peculiar velocity field
in the left, middle and right panel, respectively, as extracted from the grid around the most massive halo.  The differences between the runs that include and do not include UV
heating and gas cooling are rather large, as we expect from the
conclusions drawn from the $T-\delta$ diagrams
(Fig.~\ref{t-rho-panels}), while the differences between the $FHC$ and
$HC$ runs are very small.

\begin{figure*}
  \resizebox{\hsize}{!}{\includegraphics{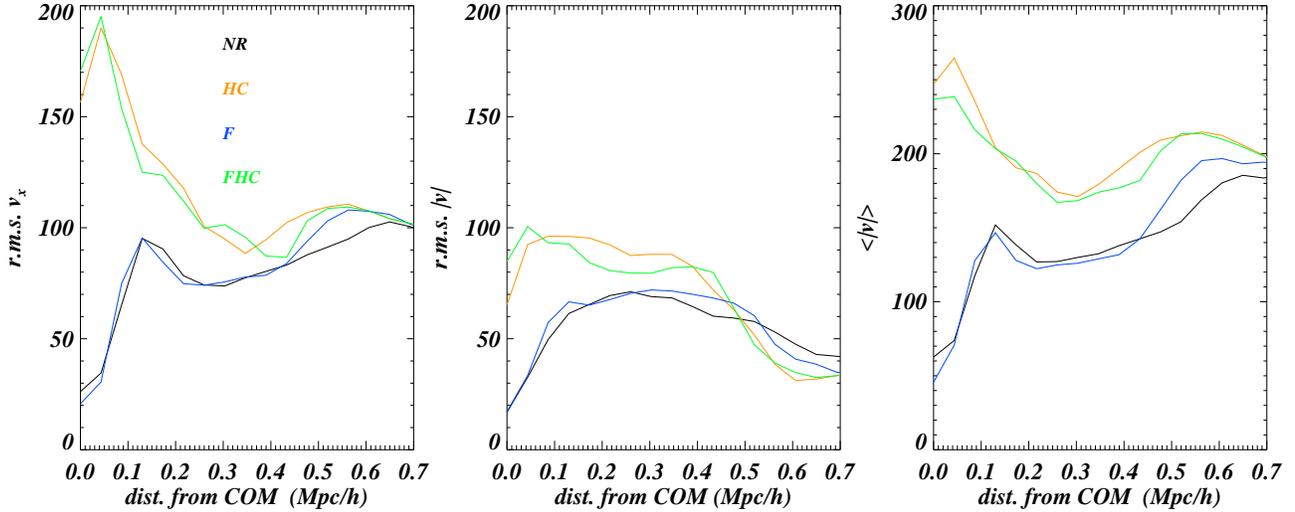}}
  \caption{Values of the IGM rms peculiar velocity along the $x$-axis
    (left panel), the rms peculiar velocity modulus (middle panel)
    and the mean of the modulus of the velocity around the most massive halo (right panel). The
    values are plotted as a function of the distance from the centre
    of mass of the halo in comoving coordinates.
    Results are shown the four different runs investigated in
    this project, colour-coded as in Fig.~\ref{halo1}.}
  \label{halo1b}
\end{figure*}

In Fig.~\ref{halo1b} we consider instead the peculiar velocity
fields and we bin the values as a function of the distance from the
centre of mass (COM) of the halo.  The velocity fields are
particularly interesting since they are expected to impact on the
properties of QSO absorption lines. In the $y$-axes we show the
rms value for the peculiar velocity along the $x$-direction
(left panel), the rms value for the modulus of the peculiar
velocity, in the middle panel) and the mean value for the
modulus of the peculiar velocity in the right-most panel, all in ${\rm km\ s^{-1}}$.

We decided to consider three different physical quantities (that are
of course related) in order not only to highlight the effects along
the line-of-sight (root mean square values for the peculiar velocity
along the $x$-direction, a quantity that is related to the broadening of the
absorption lines), but also to consider physical three-dimensional
quantities.

One can see that the there are only small differences between the four
models at distances larger than $0.5\ {\rm Mpc}\ h^{-1}$. The $F$
model results in a slightly larger rms values for the peculiar
velocity along the $x$-axis and for the mean modulus of the velocity
as compared to the $NR$ run, but these differences are at the 20 per cent
level.  At distances below $0.5\ {\rm Mpc}\ h^{-1}$ the $FHC$ and $HC$
models are very similar, and radically different from the runs without
cooling, because of the larger velocity values, which can be noticed
also in the right panel of Fig.~\ref{halo1b}. In the next Section we
will discuss possible causes for this more vigorous flow, found in the
vicinity of haloes in runs using UV heating and radiative cooling.

\section{Discussion and conclusions}
\label{discussion}

In this work, grid-based hydrodynamical simulations have been used for
the study of the features of turbulence at $z = 2$ in the medium
around galaxy-sized haloes. In the four runs considered in this paper,
we checked the effect of including the additional physics that is
required for properly modelling the circum-galactic medium (UV heating
background, radiative cooling) and, for the first time in this kind of
simulations, a SGS model for unresolved turbulence
\citep{mis09}. Turbulence is analysed at subgrid length scales, namely
just below the spatial resolution of $25\ {\rm kpc}\ h^{-1}$.

The analysis has been focused mainly on the diffuse gas component,
with baryon overdensity in the range $1 < \delta < 50$ (IGM), and on
the warm-hot phase (WHIM) with $10^5 < T < 10^7\ {\rm K}$. The gas in
the IGM phase has a lower turbulence level (Fig.~\ref{pratio-1d}) than
the WHIM. The trend with overdensity is also different: turbulence
diagnostics in the WHIM are roughly constant over the whole
overdensity range, whereas in the IGM there is a steady increase with
overdensity. For the run $FHC$, the ratio of SGS turbulent to thermal
pressure $p_{\rm t} / p_{\rm therm}$ grows from 0.002 to 0.08 in the
IGM, and has a typical value of 0.1 in the WHIM. A similar trend has
been found for a related quantity, the SGS turbulent to thermal
Doppler parameter ratio $b_{\rm t} / b_{\rm therm}$, which has a
volume-weighted average of 0.034 in the IGM and 0.26 in the WHIM,
corresponding to values for $b_{\rm t}$ of $1.02\ {\rm km\ s^{-1}}$ in
the IGM and $18.2\ {\rm km\ s^{-1}}$ in the WHIM. In both phases,
large values can be reached locally (cf.~Fig.~\ref{pratio-2d}). In the
IGM, the increase of the ratio $b_{\rm t} / b_{\rm therm}$ with
overdensity is fitted by the function $b_{\rm t}/b_{\rm
  therm} = 0.023 \times \delta^{0.58}$.

Modelling unresolved turbulence in the WHIM by means of the SGS model
may lead either to a heating of the gas, because of the dissipation of
SGS turbulence to internal energy, or to a cooling effect, because of
the buffer effect introduced in the energy budget by the unresolved
kinetic energy (cf.~Fig.~\ref{t-rho-1d}). Examples of the two cases are presented in Section
\ref{haloes}, where we show the location of turbulence production
(post-shock regions and shear flows around filamentary inflows),
together with the effect of modelling the unresolved turbulence. 

The mass-weighted average value of SGS turbulence energy in the WHIM
is $\langle e_{\rm t} \rangle = 1.10 \times 10^{13}\ {\rm erg\
  g^{-1}}$ for the run $FHC$, and is larger than the value found in
the run $F$ ($5.7 \times 10^{12}\ {\rm erg\ g^{-1}}$). Stirring is
therefore more effective when cooling and heating from UV background
are included. The analysis on the resolved velocity field performed in
Section \ref{los} further supports this result. A possible explanation
is that radiative cooling increases the density contrast of subclumps,
which with their motion drive more vigorous turbulence in the medium.

With respect to the IGM, the WHIM is more turbulent because is it
mostly located in the vicinity of forming haloes, or in the inflowing
filaments (Fig.~\ref{proj-phases}), and in our simulations turbulence
is driven only by the formation of cosmic structure through merger or
shocks. We notice that \citet{dco01} already stated that the evolution
of the WHIM is mainly driven by structure formation, with cooling and
galactic feedback playing a less important role. It is thus reasonable
to assume that the level of turbulence found in the simulations
presented here is a lower limit for turbulence in the gas around
galaxies, which other stirring sources can increase, even
substantially.

At the present stage of this work, it is hard to directly compare the
results with other observational and theoretical investigations. The
velocity shear of about $70\ {\rm km\ s^{-1}}$ found by \citet{rsb01}
compares well, as order of magnitude, with the turbulent Doppler
broadening at $\delta \sim 100$ in Fig.~\ref{pratio-1d}, but the
length scale of our analysis is somewhat larger. Good agreement can
also be found with the level of sub-resolution turbulence applied by
\citet{od09}, although their analysis refers to low redshift ($z =
0.25$).

Among the limitations of the performed approach, certainly the most
relevant one is the lack of gas exchange mechanisms with the
galaxies. First tests (Section \ref{feedback}) have shown the importance of metal enrichment, with respect to thermal feedback. Interestingly, from these tests we deduce that modelling galactic winds can affect our results in a twofold way, namely by stirring the gas and also by spreading metals in the gas surrounding haloes.

A logical follow-up of this project is, therefore, to treat galactic winds, injection of
metals and turbulence within the same theoretical framework\footnote{At smaller scales, some interesting ideas have been recently proposed by \citet{s13}, who interpreted galaxy outflows as products of unstable turbulent support in the interstellar medium.}. When modelling turbulence within this 
scheme, an advantage comes not only from the SGS model, but also from the
use of a grid-based code, which is better suited than SPH schemes for
the treatment of mixing processes (cf.~\citealt{odk12}). Of course, a
computational strategy focused also on better resolving haloes has to
be designed (cf.~\citealt{hbs13}, who focus on the gas properties around a single halo in AMR simulations). It will be also interesting to check the current results
against improved versions of the turbulence SGS model \citep{sf11},
more suitable for flows with a moderate turbulent Mach number, like
the WHIM.

\section*{acknowledgements}
The numerical simulations were carried out on the SGI Altix 4700 {\it
  HLRB-II} of the Leibniz Computing Centre in Garching (Germany), and
the analysis was mainly performed on the Intel Xeon {\it SuperMIG}
system. The {\sc enzo} code is the product of a collaborative effort
of scientists at many universities and US national laboratories. Most
of the data analysis was performed using the {\tt yt} toolkit
\citep{tso11}. L.I.~acknowledges the support received from the
European Science Foundation (ESF) for a short visit grant at the
University of Trieste, within the framework of the activity entitled
{\it Computational Astrophysics and Cosmology}. We thank J.~Regan
for technical help. 
M.V.~and S.B.~are supported by the PD51/INFN grant, by the PRIN/INAF-2009
grant ``Toward an Italian Network for Computational Cosmology'', by
the PRIN/MIUR-2009 grant ``Tracing the Growth of Cosmic Structures''
and by the European Commission FP7 ITN ``CosmoComp'' grant. M.V.~is also
supported by the European Commission FP7 ERC ``CosmoIGM'' grant.

\bibliography{cluster-index}
\bibliographystyle{bibtex/mn-web}
\bsp

\label{lastpage}

\end{document}